\documentclass[12pt]{article}
\def\ben{\begin{enumerate}}  \def\een{\end{enumerate}}
\def\beq{\begin{equation}}   \def\eeq{\end{equation}}
\def\bea{\begin{eqnarray}}  \def\eea{\end{eqnarray}}

\def\noi{\noindent}

\def\lsim{\raise0.3ex\hbox{$<$\kern-0.75em\raise-1.1ex\hbox{$\sim$}}}
\def\gsim{\raise0.3ex\hbox{$>$\kern-0.75em\raise-1.1ex\hbox{$\sim$}}}

\begin{document}

\vbox to 1 truecm {}
\begin{center}
{\bf CONSCIOUSNESS AND THE WIGNER'S FRIEND PROBLEM} \par \vskip 1 truecm

{\bf Bernard d'Espagnat}

{\it Laboratoire de Physique Th\'eorique}\footnote{Unit\'e Mixte de
Recherche (CNRS) UMR 8627}\\
{\it  Universit\'e de Paris XI, B\^atiment
210, 91405 Orsay, Cedex, France}
\end{center}
\vskip 1.5 truecm

 It is generally agreed that decoherence theory is, if not a
complete answer, at least a great step forward towards a solution of
the quantum measurement problem. It is shown here however that in the
cases in which a sentient being is explicitly assumed to take
cognizance of the outcome the reasons we have for judging this way are
not totally consistent, so that the question has to be considered anew.
It is pointed out that the way the Broglie-Bohm model solves the riddle
suggests a possible clue, consisting in assuming that even very simple
systems may have some sort of a proto-consciousness, but that their
``internal states of consciousness'' are not predictive. It is, next,
easily shown that if we imagine the systems get larger, in virtue of
decoherence their internal states of consciousness progressively gain
in predictive value. So that, for macro-systems, they may be identified
(in practice) with the predictive states of consciousness on which we
ground our observational predictions. The possibilities of carrying
over this idea to standard quantum mechanics are then investigated.
Conditions of conceptual consistency are considered and found rather
strict, and, finally, two solutions emerge, differing conceptually very
much from one another but in both of which the, possibly
non-predictive, generalized internal states of consciousness play a
crucial role.

\par \vskip 1 truecm

\noi Key words: measurement, decoherence, reality, consciousness, time.

\newpage
\noi {\bf 1. INTRODUCTION}\vskip 5 truemm

The central claim, in this paper, is that the
Schr\"odinger-cat -- or Wigner's friend~-- paradox cannot be really solved
without going deeply into a most basic question, namely: are we able to
describe things as they really are or should we rest content with
describing our experience? A priori, of course, we have a hope of doing
both at once. We think that, by enlarging our experience, and reasoning
on it, we shall progressively lift the veil of appearance and attain
knowledge of reality. But historically things did not take this turn.
It is well known that the attempts at imparting an ontological
interpretation to modern physics, and particularly quantum physics,
have met, and still meet, with difficulties. Admittedly, these
obstacles are not insuperable, but still they imply that essentially ad
hoc and quite artificial-looking changes should be made in the
formalism. And this is one of the reasons why the alternative
standpoint, centered on the description of communicable human
experience, is also considered reasonable after all. Of course another
reason -- a historical one -- is the fact that such a standpoint was
taken up both by the ``founding fathers'' of quantum physics and, in the
same period, by the Vienna Circle positivists. Unfortunately, it must
be observed that, on the whole, neither the ones nor the others were
explicit concerning the very existence of the dilemma in hand. Schlick,
for example \cite{1r} had clearly stated the basic positivist axiom that the
meaning of a scientific statement {\it boils down} to its method of
verification, which obviously makes the notion of ``mind'' (the mind or
minds that verify) prior to that of scientific reality. But still, in
spite of this, in most of their writings the logical positivists
implicitly seemed to suggest that the rules they claimed science should
follow (the principle of verification for example) somehow led to some
knowledge of a reality ``out there'', in other words of mind-independent
reality. \par

This ambiguity got transferred to quantum physics and, in
fact, it is still with us. In a sense it should be considered
beneficial. Thanks to it, most physicists do not think it necessary to
take sides concerning two opposite and equally unpleasant views, the --
seemingly aberrant -- one that mind is actually the ``basic stuff'' and
the technically disturbing one that words such as ``observables'',
``measurements'' and so forth should be banned from basic physics... Which
is fortunate since it turns out that explicitly opting in favor of one
of these standpoints is not at all necessary for doing research in that
science. In it, the vague notion of ``empirical reality'' serves as a
conceptual basis and is, for practically all purposes, a totally
sufficient one. On the other hand however, it is a general rule of
reasoning that, wherever they appear, ambiguities should be removed.
And sometimes this indeed is even useful in practice. I claim that this
is the case concerning the problem in hand and that, for solving it,
the rule in question should be followed, even at the price of having to
openly face one or both of the two just mentioned queer views. Here
both of them will therefore be taken seriously. In fact, it will be
shown that either of them constitutes a suitable framework for solving
the Wigner's friend paradox, provided that they are taken with all the
consequences each one implies, disregarding their incompatibility with
such and such deeply engrained ``received views''.\\

In view of the foregoing it would not be entirely unreasonable to
consider that going into such considerations somehow amounts to
deviating from physics proper. It is therefore not surprising that most
physicists turn away from such problems. But still, this disinclination
is far from being fully general. In our times some first rate
physicists do take great interest in the question of how quantum
mechanics is to be interpreted, whether or not it is compatible with
realism, and so on. And Asher Peres is distinctly one among them, as
the number of papers in which he touched upon such matters convincingly
shows. It is therefore a pleasure to dedicate this article to him. It
is true that, in such a field, most physicists -- Peres included~--
manage so as to avoid speculating, which renders their statements
concise and, correspondingly, leaves a few questions open. Here I shall
not be so careful. I shall quite avowedly -- but, still, not wildly! --
speculate. \par

The paper is divided into four parts. In the first one
(Section 2) general facts are perused concerning the measurement
problem and the extent is discussed to which decoherence theory may be
considered to solve the latter, particularly in the case in which,
besides instruments, the measurement process is explicitly assumed to
involve also conscious beings.  The second part (Section 3) describes a
tentative solution to the just mentioned problem -- often called the
``Wigner's friend problem''~-- based on an explicitly ontologically
interpretable approach to quantum mechanics (in fact, on the
Broglie-Bohm model). The third part (Section  4) explores the
possibilities of building up a solution in line with the general ideas
underlying the just mentioned one but freed from the condition of
ontological interpretability, hence essentially compatible with the
general philosophy of standard quantum mechanics. It deals with
criticisms of a logical-conceptual nature that might conceivably be
raised against it, and shows they can be overcome. Finally, in the
fourth part (Section 5) possible bearings of this approach on an
age-old philosophical problem, the one of the nature of time, are
briefly sketched.\par \vskip 1 truecm

\noi {\bf 2. DECOHERENCE AND MEASUREMENT}\par \vskip 5 truemm

As is well known, the main difficulty
with quantum measurement theory is that when the system $Q$ on which a
quantity $B$ is to be measured (by means of an appropriate instrument) is
not, initially, in an eigenstate of $B$, if the instrument is described
quantum mechanically the Schr\"odinger time evolution leads, for the
overall system composed of $Q$ and the pointer (or of $Q$ and the rest of
the world if, along with the pointer, we take the environment into
account, as we should), to a state that is a superposition of
macroscopically distinct states or a combination thereof. This result
seems hardly compatible with the often held view that the basic quantum
mechanical symbols describe reality as it really is quite independently
of us, since what is observed has no clear relationship with such
superpositions. And, what is more, under the assumption that the
quantum predictive rules are universal the superposition in question
could be shown \cite{2r,3r} to be incompatible with the view that macroscopic
objects always have definite localizations.\par

 Then decoherence theory came
in. As we all know, it is grounded on the fact that all macroscopic
systems significantly interact with their environment (including their
``internal'' one) and on the remark that, in practice, most of the
physical quantities pertaining to the environment cannot be measured.
It is claimed by many that decoherence actually solves the measurement
problem. This assertion however is far from being endorsed by all
physicists, and the reason is that recognition of the universality of
the said interaction is only one of the ingredients in the solution.
Another one is a watering down of realism. \par

This point being quite
crucial for what follows, some details are worth being recalled
concerning it. In quantum mechanics it is usually and appropriately
considered that the meaning of factual statements is directly tied to
what we count as evidence for them. More precisely: in it a factual
statement, if true at all, can be true only in virtue of something of
which we could know. In particular, statements concerning the physical
state in which a physical system lies can have truth-values (be true or
false) only in virtue of measurements that we could perform. Hence it
is only by referring to the possible outcomes of the measurements of
some observables pertaining to a physical system that we may define the
state of the latter. A strict, somewhat Schlick-like, interpretation of
these epistemological principles leads to the (strong) completeness
hypothesis, according to which, referring to the measurement outcomes
of a complete set of compatible observables entirely specifies the
physical state in question. This argument corroborates the commonly
accepted view that states of individual systems are specified by state
vectors\footnote{It is true that the notion of ``protective
measurements'' (in which the interaction Hamiltonian acts for a long
time with low intensity) makes it possible, in principle, to impart an
operationally defined meaning to mean values of observables of
individual systems, and hence also to the notion of density matrices
attached to such systems \cite{4r}. To my knowledge the question whether this
new idea might possibly serve as a basis for new attempts towards a
realist approach of the measurement problem has not been examined.} .
\par

It is on this basis that, in one of the above referred to
papers, Bassi and Girardi \cite{3r} could build up quite a general proof that
decoherence fails to solve the measurement problem. They proceeded as
follows. Given the macroscopic configuration of a macro-object of any
sort they considered the set $V$ of all the state vectors that may
represent it. Next, they pointed out that, assuming the (strong)
completeness hypothesis, the sets of vectors corresponding to two
well-separated macroscopic configurations should be ``almost orthogonal''
in a mathematically well-defined sense. In particular this must hold
true concerning the sets, $V_U$ and $V_D$, that, in a measurement process,
contain the final overall state-vectors corresponding to two distinct
values, $U$ and $D$, say, of the measured quantity. They then considered a
generalized such process, such as the above-described one, in which the
to-be-measured quantity $B$ may have one of the two values $U$ and $D$.  And
they could show that, in virtue of the quantum evolution law, the final
state-vector of the overall system (including the instrument pointer as
well as the environment) can then be neither in $V_U$ nor in $V_D$. (nor
indeed in any other macroscopic position different from ``$U$'' and ``$D$'').\par

In view of this, an upholder of the view that pointers always {\it are} at
definite places (in short, a ``realist'') may not consider that
decoherence solves the measurement riddle. This shows that for
decoherence to be significant a further move, one of a philosophical
nature, is necessary. Roughly speaking it consists in taking two
``conceptual steps'' successively. The first one is to consider, along
the lines marked out by Plato, Descartes, Kant and others, that our
senses may, to a great extent, be deceitful and that what we apprehend
-- the set of the phenomena -- is liable to considerably differ from
whatever may be said to ``really exist''.  Such a view makes it easier
for us to grant that, after all, there might be a gap between what we
see -- pointers positions -- and what we think of -- state vectors. On the
other hand, the view in question is still a purely negative one. It
does not positively tell us what type of information mathematical
entities such as state vectors may provide us with. Our second step
must be a positive one in this direction. And it seems that there is no
other one to be taken than just to turn to the (generalized) Born rule,
otherwise said to the rule yielding the probabilities that, upon
measurement of a physical quantity, such or such value should be
obtained. Now, what is most important at this point is that the rule in
question is essentially {\it predictive of observations}. It has no
ontological significance. It does not describe objects and their
properties (it is well known that attributing such a function to the
Born rule would immediately raise a host of conceptual difficulties).
It merely informs us of what we shall see -- or of the chances we have
of seeing this or that -- if we perform such and such actions. \par

Hence
there are some good grounds for considering quantum theory to be
essentially -- and exclusively -- predictive of observations. Now, it is
true that if quantum mechanics is considered universal this limitation
partly deprives physics of its ``explanatory power at first level'',
grounded on the notion of physical events being explanatory causes of
other physical events. It implies that ultimately such notions as those
of cause and explanation should be either dropped or (as seems more
rationally justified \cite{5r}) transferred to the ``higher realm'' of some
``veiled reality'' lying outside the direct reach of science proper. It
is therefore understandable that some of us should consider the
limitation in question to be a pity. But it may be claimed that, in a
strict scientific sense, it is not. After all, describing ultimate
reality as it really is always was the role taken up by metaphysics,
and it was often claimed that the success of science was due to just
its parting with metaphysics. In this respect there is therefore much
truth in Schlicks \cite{1r} above cited axiom according to which the meaning
of a statement is nothing else than its method of verification, which
is, indeed, tantamount to saying that statements that look descriptive
are, in truth merely predictive of observations.\par

 It follows that the
fact quantum mechanical predictive methods are so good as to never have
been found at variance with experimental tests may legitimately be felt
to remove our conceptual qualms relative to this theory, including
those concerning measurement. And this is especially true since it may
then legitimately be claimed \cite{6r} that, at least when only inanimate
instruments are involved (see below), decoherence solves the
measurement problem. Basically this is due to the fact that, to repeat,
if Schlick's views are taken literally, when we say that ``obviously'',
in an ensemble $E$ of systems $S$ that include pointers, each individual
pointer is in one definite scale interval, the word ``is'' should not be
misunderstood. In fact, it does not describe a state of things. It
merely means that if we look at $E$ we shall have the impression of
seeing the pointers distributed as just said. True, we also have
something else in mind. Accustomed as we are to the ``descriptive'' sense
of the verb ``to be'', by using this verb here we also express our
expectation that, of all the predictions concerning practically
feasible measurements that are normally inferred from its use, none
will be at variance with the data. Fortunately, in virtue of
decoherence this condition is fulfilled as we know.\par

 It is important to
observe that this ``predictive'' or ``purely operational'' interpretation
of quantum mechanics happens to remove two well-known conceptual
difficulties that, within the more realist interpretations, still beset
the decoherence theory of measurement. The first one is tightly
connected with what was just noted. It is known that, concerning some
extraordinarily complex measurements also involving the environment,
quantum mechanics unavoidably yields probabilities that are at variance
with consequences of our usual ``realist'' way of picturing ``outside
macroscopic reality''. But here, and especially since (as has now been
shown \cite{7r}) the predictive rules of classical mechanics are mere
consequences of the quantum mechanical ones, we may completely drop any
realist picture. We may center on observational predictions. And it is
then clear that the discrepancy in question, exclusively bearing as it
does on practically unfeasible measurements, is void of both
theoretical and experimental significance. \par

The second difficulty we
have in mind is the famous ``and-or'' one. Within the more current (and
more ``realist'') approaches, applying decoherence theory to individual
cases rather than ensembles raises quite serious problems. Let it be
stressed that such a difficulty does not actually arise here. The
reason is that, to repeat, quantum mechanics is now viewed as merely
being a set of computational rules informing us, via the (generalized)
Born formula, of the probabilities we have of observing this or that.
And it is in the spirit of the approach in hand that these rules,
including the Born one of course, should be considered truly primitive.
As such, they do not have to be explained, or inferred from anything
else. They are just given as they are. And in particular, the Born rule
is not derived from ensemble considerations... It is given in its
probabilistic form, that is, it yields the probabilities that we shall
have such or such specific, ``individual'' impression. Admittedly,
experimentally verifying the correctness of the thus computed
probabilities necessitates considering ensembles. But this was already
the case, in classical physics, concerning all the theories that
involved probabilities. In other words the probabilities that, making
use of decoherence, we calculate on the basis of the quantum rules
apply to individual events in the same way as do those a card player
attributes to the event of pulling out a given card.\par

So, upholders of the views of Schlick and the most thoughtful authors
of philosophy of science books may well consider that everything is
quite in order. On the other hand, considered under this light quantum
mechanics obviously has the notion of consciousness, not the notion of
reality, as a referent. It deals, not with what is but with what we
perceive. More precisely, it involves two basic notions, one having to
do with actuality and the other one with potentiality, both related to
consciousness. The ``actual'' one is that of ``state of consciousness'' or
``state of mind'' (distinguishing these two notions will not be here
necessary): When a human being has made some observation it is quantum
mechanically meaningful to state that he or she ``is'' in the
corresponding state of consciousness. As for the ``potential'' notion, it
is the one of ``probabilities of observation'' following from applying
the quantum rules to the pieces of knowledge the state of consciousness
contains. \par

Now, it so happens that this state of affairs raises new
questions. The point is as follows. True, quantum mechanics does not
provide us with anything like a God's Eye view on what {\it is}. That much we
just saw. But still, in fact it gives us -- or seems to give us -- a
point of view that somewhat partakes of a God's Eye one. It does so in
the sense that the predictions quantum mechanics makes are held to be
true for {\it a whole abstract community of human beings}. For, in fact, a
kind of disembodied Experimentalist, who is supposed to be looking at
an ensemble of measurement-performing instruments and whom these
predictions inform about the proportion of the latter on which he or
she will see the pointer at such and such a place on the dial. This
would raise no problem if only one Experimentalist, or, say, one
conscious being existed. Unfortunately (as Wigner quite appropriately
remarked!) there are several ones. Unavoidably we therefore have to
face the so-called ``Wigner's friend problem'', which is just the
Schršdinger-cat one, only, with a cat explicitly endowed with
consciousness. \par

This raises, as I said, a question. To explain what it
consists of, let us consider again a ``measurement process'' (in the
generalized sense) bearing on a microscopic quantum system $Q$ and let us
assume that a physicist $P$ has prepared $Q$ in a state that is not an
eigenstate of the to-be-measured observable $B$.  Let then $S$ be the
system composed of $Q$, the instrument and a friend $F$ of $P$ looking at the
pointer. Let us think of $S$ at a time $t$ when the
process is over but $P$ has not yet looked at the result (nor asked $F$
what he saw). Within a statistical ensemble $E$ of such $S$'s, each one of
the friends then sees ``his or her'' pointer lying in a graduation
interval corresponding to one definite eigenvalue of $B$.  From this
actual knowledge, and assuming quantum mechanics is exactly true,
he/she may make definite (probabilistic) predictions concerning what
results would be obtained in the future if such or such measurements
were done. All these predictions may then be combined according to
classical, standard probability rules, so as to yield the probabilities
with which $P$ should herself obtain these results. As we know, there are
physical quantities (involving the environment) concerning which these probabilities do not
coincide with those $P$ directly obtains by means of a quantum-mechanical
calculation based on the content of her own state of consciousness,
{\it without} assuming the friends to be in definite consciousness states.
True these quantities are, in virtue of decoherence etc., not measurable in
practice. But, in the present context, this remark does not remove the
difficulty. The reason is that, in fact, we here have to do with a
question of logical consistency. Assuming both that quantum mechanics
is a universal theory and that the states of consciousness of the
friends are predictive same as the one of $P$ is (i.e. assuming the
predictive quantum rules may be applied to them same as to it) would
amount to putting forward a theory yielding different results according
to the way calculations are made. This is unacceptable, quite
independently of whether or not the results are practically checkable.\par

Note that this objection cannot be raised against the general argument
on the basis of which decoherence is said to ``solve'' the measurement
problem (when the consciousness of the ``friends'' is not brought into
the picture). The reason is that, to repeat, within the conception in
hand the only data that lie in the realm of actuality (i.e. to which
the verbs ``to be'' or ``to have'' may, strictly speaking, be applied) are
states of consciousness. Now, in the just mentioned case, the only
relevant state of consciousness is that of $P$, which merely contains
potentialities of observations, informing her of the chances she would
have of getting this or that result if she chose to measure such and
such observable. True, $P$ likes to think of these results as referring
to some empirical reality. She expects that, of all the predictions
concerning practically feasible measurements that may be inferred from
assuming this to be true, none will be at variance with what she
directly derives from what she knows, without making that assumption.
But, to repeat, decoherence provides her with proper insurance in this
respect. Within the here taken up conception of what quantum mechanics
is and describes it is therefore true that the above objection is
restricted to the case in which conscious ``friends'' are involved.\par \vskip 1 truecm

\noi {\bf 3. THE ``WIGNER'S FRIEND'' PROBLEM, A MODEL}\par \vskip 5 truemm

The just mentioned conception may be referred to
as the ``purely operationalist'' one\footnote{The content of this
Section appeared in a preliminary form in Ref. \cite{8r}}. But
note that it does not in any way amount to rejecting -- as meaningless --
the notion of a basic mind-independent reality, as pure idealist
thinkers do. It consists in observing that, whatever this entity
``really is'', it presumably differs even more than previously thought
from what it looks like. And that, consequently, the more secure
standpoint is not to take sides, in an a priori manner, on the question
whether or not it should be pictured and, if it should, how it should
be. Observe in this respect that within this standpoint the
completeness assumption should not be stated in the standard, strong
form: ``hidden variables do not exist'' that, following Bassi and
Ghirardi, we implicitly imparted to it in Section 2, since this is a
metaphysical hypothesis, even though a negative one. Following Stapp
\cite{9r}, the assumption in question should be expressed as the hypothesis
that ``no theoretical construction can yield experimentally verifiable
predictions about atomic phenomena that cannot be extracted from a
quantum theoretical description''. This leaves open the possibility that
hidden variables exist, provided that they should be ``really hidden''.
Otherwise said the operationalist conception is, as here understood,
quite general. For example, it is compatible with both pure antirealism
and Bohm's ideas, especially when the latter are expressed via the
conception of a real {\it implicit order}, differing very much from the
{\it explicit} one that reflects but the appearance of mind-independent
reality. \par

Now, it turns out that the latter remark is here of help, in
that it provides us with some sort of a guiding line. For indeed, when
faced with riddles such as the one described in Section 2 we,
physicists, feel somewhat at a loss until we find some similarity
between them and problems that are, to us, more familiar. In this
respect, the old Louis de Broglie-Bohm (hereafter B.B.) hidden variable
model (with pilot wave or quantum potential) may be useful. Not that we
should necessarily believe it is true. Many arguments (e.g. those
described in detail in \cite{5r} and \cite{10r}) speak against it. But it does
reproduce all the observational quantum mechanical predictions; it
yields, to the quantum measurement problem, a solution differing from
the above one but fully consistent as well\footnote{In it, the representative point is, right from the start,
determined to proceed  into one or the other of the sectors of
configuration space corresponding to the possible pointer positions.
The reason why this solution is not at variance with the Bassi and
Ghirardi proof is, as pointed out by these authors, that, in this
model, the strong form of the completeness assumption is not assumed.}; and it has the
great advantage of being conceptually crystal-clear. It can therefore be used
as a ``theoretical laboratory''. If, within it, it proves possible to
take explicitly into account the (unquestionable!) fact that Wigner's
friend is conscious, it is conceivable that the basic idea underlying
this solution can be extended, outside the model, to the general
theory. Incidentally, note that in the model the above-mentioned
difference between implicit and explicit orders of course holds. The
implicit order concerns the hidden variables that, together with the
nonlocal pilot-wave, compose {\it mind-independent reality}. The explicit
order is the order that is manifest in the appearances that compose the
set of the observed phenomena, alias {\it empirical reality}. \par

According to
the B.B. model, within a Young-type thought-experiment with two slits
the particle is driven by a pilot-wave that passes through both slits
at once; and this has the consequence that, in the model, while
each particle passes through but one slit, fringes nevertheless appear.
We can therefore say that, in the B.B. model, the particle is at any time at
some well-defined place even though it takes part in a typically
quantum phenomenon. In this respect it resembles the friend in Wigner's
apologue, who is at any time in some well-defined state of
consciousness while he also is taking part in a quantum phenomenon. To
strengthen the analogy it is then appropriate that, in the model, we
should attribute to the particle some kind of a mentality (or, say,
proto-mentality), the physical nature of which needs not be specified
in detail. Within the experiment in question, each one of the involved
particles then ``observes'' that, at a certain time, it passes through
one, well specified, slit. This is an {\it internal state of consciousness}
of the particle and since, in the model, the particle position is an
element of mind-independent reality, this internal state of
consciousness should also be considered as being an element of
mind-independent reality. For the particle, this internal state of
consciousness has no predictive power, since what will happen to the
said particle is entirely governed by the pilot wave. \par

Now, it is often,
and quite rightly, said that the very fact of knowing through which
slit the particle passes prevents the fringes from being formed. At
first sight this might seem to constitute a valid objection against any
idea of attributing a state of consciousness to the particles. And in
fact, so it would be if we assumed that the particle could communicate
its knowledge to the world at large. So let us assume that it can't.
That, at least as far as micro-systems are concerned, ``{\it internal states
of consciousness}'' are really private (since hidden variables do not act
on the pilot wave, such an idea is quite consistent with the fact that,
in the model, the ``consciousness state'' in question is a ``hidden
variable'' just as the position itself). This being the case, an
external observer such as our $P$ above, even if we assume she knows of
the {\it existence} of such internal states of consciousness, must explicitly
ignore this existence when predicting what will be observed. She
therefore predicts the fringes will appear. And this prediction agrees,
as we know, with experiment. \par

On the other hand, imagining a category of
states of consciousness that always remain totally hidden would
obviously be quite pointless. Hence, we should ask whether
circumstances exist in which statements bearing on such ``internal''
states may, after all, have some relationship with the public domain of
shared experience (while remembering of course that this domain is the
one of the {\it explicit} order, that is, in the model, the one of
``appearances that are the same for everybody''). Now, decoherence helps
us here. For, in the Young-type experiment, call $\phi_1$ and $\phi_2$ the partial
wave functions issuing from slits 1 and 2 respectively, and suppose we
replace the micro-particles by corpuscles that are appreciably larger
and whose interaction with the environment is, consequently, not
negligible. The fringes then fade and, when the corpuscles are
macroscopic enough, they practically disappear. For the purpose of
predicting outcomes of future observations, the ensemble of the
involved corpuscles may then be treated as a mixture of two ``pure
cases'' described by the wave functions $\phi_1$ and $\phi_2$. Now $\phi_1$ ($\phi_2$) is
just the wave function that an observer would attribute to a set of
corpuscles known to have passed through slit 1 (2). This shows that, in
such circumstances, the internal state of consciousness of the
corpuscles passing through one particular slit may indeed be considered
without harm to have the intersubjective predictive role normally
attributed, in quantum mechanics, to pieces of knowledge obtained from
measurements. Incidentally, note that this reasoning is fully
consistent with the Broglie-Bohm model since, as far as mere
predictions are concerned, this model yields the same ones as
non-relativistic quantum mechanics. \par

If we now turn back to the Wigner's
friend problem and still consider it within the Broglie-Bohm model, we
find that the foregoing views can quite naturally be fitted to it.
Indeed, for the same reasons as above, we may assume without
inconsistency that $S$ has an internal state of consciousness as long as
we assume it has no predictive power. And we can also relax somewhat
the latter assumption when we take into account the fact that $S$ is
macroscopic and interacts therefore with its environment. In fact, for
the purpose of predicting what will {\it practically} be observed,
(forgetting about measurements that are conceivable only in principle)
an ensemble $E$ of thus prepared $S$'s can be treated as a mixture, for
decoherence is at work. And -- just as above -- the quantum states
composing this mixture are the ones that the various possible internal
states of consciousness of the friends would generate if these
consciousness states were viewed as predictive. Hence these internal
states of consciousness, which, related as they are to hidden
variables, are basically ontological, still may be considered to also
represent elements of {\it empirical} reality. More precisely they can be
viewed as coinciding in nature with the predictive states of
consciousness we normally refer to when we state that such and such a
measurement outcome has been observed.\par

 To sum up, within this
conception (or ``model'') it is considered that even microsystems can be
endowed with ``internal states of consciousness'' (or
``proto-consciousness'', whatever this may be) that are elements of a
basic, not publicly accessible, reality, rather than of empirical
reality. In other words, they are hidden (remember we left open the
possibility that hidden variables should exist, provided that they
should be ``really hidden''). It is only when the involved systems become
macroscopic enough for their interaction with the environment to be
appreciable that these internal states of consciousness obtain some
degree of public significance. This means that they gain predictive
power. More precisely (as we easily realize by thinking of intermediate
cases in which ``not quite macroscopic'' systems are involved) they make
it possible to correctly predict the outcomes of a certain class -- call
it $A$ -- of observations whereas they yield incorrect ones concerning
those of another class -- call it $B$.  Now, it is a fact that (due to the
nature of these two classes) human beings perform observations of class
$A$ much more easily than observations of class $B$.  And indeed this is
true to such an extent that when the involved systems are thoroughly
macroscopic, observations of class $B$ are, as a rule, practically
unfeasible, as we know. Moreover, it is also the case that the
impressions corresponding to the outcomes of measurements of class $A$
may usually be described in a realist {\it language}, that is, {\it as if} they
referred to objects existing per se. The set of such intersubjective
appearances is what is called here ``empirical reality''. It is thus
meaningful to speak of a kind of ``co-emergence'' of, on the one side,
``public'', states of consciousness that are {\it practically} predictive
(although, concerning class $B$ observations, they are not), and, on the
other side, empirical reality.  In line with one of the foregoing
remarks this co-emergence is to be thought of as (a-temporally) taking
place out of a ``mind-independent reality'' that, itself, presumably lies
beyond our intersubjective abilities at describing. \par

It is interesting
to note that, surprisingly enough, this model shows a similarity with
Whitehead's views. For indeed the notion of internal states of
consciousness is quite akin to the ones of ``prehensions'' and ``occasions
of experience'' that, in Whitehead's philosophy, play a basic role even
at the elementary particle level.\\

\noi {\it Discussion} \par \nobreak Before proceeding further we should, as a precaution, ask
ourselves whether the model is, after all, fully consistent. The reason
why this question arises is as follows. Although the Bohm theory on
which the model rests is ontologically interpretable by construction,
still, in it, the difference between the implicit and explicit orders
is basic and unavoidable. Which, to put it bluntly, means that the
explicit order (alias empirical reality) is, in a way, but an illusion
of our senses. This raises a question since the here described model
takes macroscopic systems to be elements of mind-independent reality.
Now, by definition so to speak, macroscopic objects are localized in
definite regions of space whereas Bell's theorem shows mind-independent
reality to be non-local. At first sight it would therefore seem that
macroscopic objects couldn't be elements of the latter (alias of
implicit order). A moment reflection shows however that this objection
has no real substance. The point is that the Broglie-Bohm theory is
ontological and that, in it, complex systems are composed of corpuscles
each of which has, at any time, quite a definite localization in space.
In virtue of the interaction existing between them, many such
corpuscles may then constitute a stable localized composite system.
This does not violate the Bell theorem since, in the B.B. model,
non-locality essentially takes the form of instantaneous interactions
that do not decrease when distance increases. Admittedly the complex
systems in question somehow interact this way with one another (through
the non-local overall wave function or, equivalently, the quantum
potential) but this in no way prevents them from existing as definite
entities. And indeed Bohm himself forcefully claimed (Bohm and Hiley
\cite{11r}, ch. 8) that, in his model, macroscopic objects do exist, not only
as elements of the ``explicit order'' but quite independently from us,
that is, in an ontological sense.\\

\noi {\it Remark}\par \nobreak
 Consider, within the B.B. model, a two photon
correlation-at-a-distance experiment, of the type used for checking
Bell's theorem, and assume that two distant observers $A$ and $B$
successively make, in this time order, polarization measurements along
one and the same direction. According to the foregoing their internal
states of consciousness are elements of mind-independent reality. On
the other hand, the results that $A$ and $B$ get are strictly correlated,
but we know from the Bell theorem that this correlation is not due to
``common causes at the source''. John Bell ([Ref.12], ch.15) explicitly
showed it follows from the fact that what takes place in $A$'s instrument
directly influences -- via the non-local pilot wave -- the behavior of
the photon in $B$'s instrument. In other words, contrary to what we
naively expect, the outcome of $B$'s measurement does not depend at all
on the hidden variables of the photon arriving to $B$.  And therefore,
since, in our model, $B$'s internal state of consciousness is strictly
linked to the hidden variables affecting $B$ we must consider the said
internal state to be non-predictive. But of course both $A$ and $B$ have a
natural tendency to take up the view according to which (i) their
internal states of consciousness {\it are} predictive and (ii) the
correlation they observe is due to common causes at the source. This
view may be considered to be the embryo of a conception of empirical
reality. In the case in hand such a conception is obviously deficient
since all the tests of Bell's inequalities violation prove it is wrong.
But when photons are replaced by somewhat more complex corpuscles, the
corresponding tests rapidly become extremely difficult to make. In fact
they quite soon become practically unfeasible. Correspondingly the
conception of empirical reality the embryo of which we just described
becomes more and more credible. Finally, when macroscopic objects are
involved in place of photons, it becomes established truth. This is all
right, after all. But it is conceptually and logically all right, only
provided we are aware that, when we express ourselves in this way, we
merely refer to an ``empirical reality'' that crucially depends on human
aptitudes. \par

Again, in this model it is legitimate to speak of a kind of
``co-emergence'' of empirical reality on the one hand and states of
consciousness on the other hand.  But it is so with the important
reservation that the said states of consciousness are not those that
are deepest in our mind (the internal ones, which alone are
ontological) but just the predictive ones, which refer but to Bohm's
``explicit order''.\par \vskip 1 truecm

\noi {\bf 4. BACK TO STANDARD QUANTUM MECHANICS} \par \vskip 5 truemm

 In Section 3 we used the
ontologically interpreted Broglie-Bohm theory as a theoretical
laboratory and we constructed a model. It is true that along with
distinct advantages, the said theory has in it quite unpleasant, well
known features. But, to repeat, our idea has been that if, within it,
it proves possible to take explicitly into account the fact that
Wigner's friend is conscious -- as we have just found is indeed the case
-- it is conceivable that the basic idea underlying this solution may be
extended to a much wider theoretical framework. It is the feasibility
of this that is now to be examined. \par

Let us first observe that there
exists a fully consistent and very simple way of doing this. It
consists in observing that the features of the B.B. model we made use
of are but general ones. In fact, they boil down to the idea that the
notion of a basic, mind-independent reality is meaningful and that the
quantum mechanical symbols do not necessarily yield the finest possible
description of it, so that the completeness assumption may be taken up
merely in the weak, Stapp, sense. Now there is, of course, no reason to
believe that the B.B. model is the only possible ontological
interpretation of the quantum observational predictive rules. Indeed,
other models are available that also have the general features in
question. Hence a fully reasonable standpoint is to assume what
follows. The notion of reality does have an ontological significance
even though we don't really know what reality consists of. Objects have
an ontological status. Minds, with also an ontological status, are
attached to some of them. And quantum mechanics happens to provide
minds, directly or indirectly, with reliable observational predictions
in, as it seems, all the various domains of physics. Within such a
conception space and time -- or space-time, or cosmic time~-- also enjoy
an ontological status. They are arenas in which quite real events take
place. Clearly, such a world-view is general enough to allow for the
possibility of developing within it considerations akin to those
unfolded in the foregoing section. It may therefore be claimed that,
from a quantum-mechanical point of view, it constitutes an acceptable
metaphysics.\par

 On the other hand, since the advent of quantum mechanics
it was always considered imperative to avoid anything resembling
mechanistic, or too realistic, models. And, to repeat, it was claimed
that to this end one should abide to the basic epistemological
principle that statements unrelated with anything we could get informed
about through appropriate measurements are meaningless. According to
this view, ontological commitments should be banned. As we saw, in
quantum mechanics following such a line of thought results in
attributing primeval importance to the Born rule. Which, since this
rule is fundamentally predictive of observations, amounts to consider
quantum mechanics to be essentially and exclusively predictive of
observations. It is therefore appropriate that we should inquire
whether or not the way of removing the Wigner's friend paradox put
forward in the foregoing section is susceptible of being transposed
into the framework of such a conception of quantum mechanics.\par

 A preliminary question arises at this point, namely: is there a risk that
this purely observational predictive nature of quantum mechanics should
jeopardize the validity of the whole decoherence-based quantum
measurement theory? In this section it will first be shown that,
concerning measurements that involve but inanimate instruments this,
fortunately, is not the case. But it will also be shown that indeed,
same as above, a problem is thereby raised as soon as animated
participants are involved. It will however be proved that, when all is
said and done, here also the problem in question may be solved by
adopting, concerning minds, the views stated in Section 3.\par

 The reason why, prima facie, we might wonder whether the standard quantum
measurement theory is as self-consistent as we usually believe it to be
is of the same general nature as the one that motivated the {\it discussion}
towards the end of Section~3~... although it ``pulls in the opposite way''
so to speak.  While the difficulty came there from the fact that
macroscopic objects seemed not to be ``real'' enough, here it comes from
them looking too ontologically real. The point is this. In Section 2
we noted that for showing that decoherence does remove the measurement
riddle we had to drop the idea that pointers {\it are intrinsically} in such
and such states corresponding to definite places on the dial. We
observed that such a descriptive view has to be replaced by an approach
purely predicting observations. But on the other hand, in order to show
that the measurement riddle is really removed, we eventually had to
take into account the fact that instruments are macroscopic. And this
was by no means just an observational prediction. It was a statement of
a descriptive sort. Now, is the occurrence, in the theory, of a
statement of such a nature compatible with the view that the said
theory should be purely predictive and in no way ``ontological''? \par

The question sounds debatable but it should be answered positively. To
require that a theory should merely be predictive of observations does
not mean that it should involve no descriptive assertions. It means
that the constitutive statements of the theory should be of the form
``in such and such circumstances we shall observe this or that'', and
this very structure implies that the circumstances in question should
be {\it described}. Now, the requirement that the theory should not be
ontological implies in turn that these circumstances should be mere
phenomena, that is, should be referred to human experience. But here
this raises no problem for in standard (non-ontological) quantum
mechanics this requirement is satisfied. The set of the quantum rules
includes the time-independent Schr\"odinger equation, which predicts what
types of objects we shall perceive; and this equation shows that among
such objects there may be bound states involving a great number of
particles. In other words, macroscopic objects, far from necessarily
belonging to the ``ontological'' realm may consistently be considered to
be mere phenomena, in a Kantian sense. Moreover, the same Schr\"odinger
equation informs us that the energy levels of such objects must lie
very close to one another, so that the said objects have non-negligible
interactions with their environment. It follows that the
decoherence-based measurement theory as it is reported on in Section 2
is indeed fully consistent with what we called the operationalist
conception of quantum mechanics.\par

 On the other hand the same cannot be said concerning the theory of measurements involving animated observers
(cats or ``friends''). As repeatedly noted above, this operationalist
conception centers on the notion of consciousness, which (apart from
basic unknowable reality, see below) indeed is, in it, (as, by the way,
in logical positivism!) the only primitive one (objects and so on
essentially being ``what is perceived by consciousness''). According to
it the whole of quantum mechanics deals with the impressions
consciousness will get under such and such circumstances. And, since
physics essentially deals with inanimate objects, in this science the
multiplicity of conscious beings does not normally constitute a
problem. As pointed out above, the predictions quantum mechanics makes
are held to be true for a whole abstract community of human beings,
otherwise said, for a kind of disembodied Experimentalist, assumed to
be looking at instruments. But then, when it is assumed that a ``friend''
(or, for what we know, just even a cat) looks at the pointer, the
situation radically changes. For, clearly, the consciousness of this
individual may not be lumped together with the one of the disembodied
Experimentalist, that is, in the present instance, with the one of the
people who started the experiment. But, on the other hand it must
obviously be considered to stand on the same ``ontological level'' as the
latter, for Wigner's friend surely is just as thoughtful a person as
you and I.  If the consciousness of the experimentalist -- call her $P$
again -- who initiated the experiment is objectively real, the one of
the ``friend'' must be real as well. Now, in the operationalist
conception this raises a problem for, as we just noted, in it the
instruments of observation are not considered to exist per se. They are
mere ``objects for us'', that is, in the case in hand they ultimately are
referred to $P$'s consciousness. They are just parts of what $P$ perceives,
otherwise said they are, in a sense, mere appearances. Hence the body
of the ``friend'' is a mere ``appearance to $P$'' as well. And it seems
absurd to assume that his consciousness, which, as we just pointed out,
is just as real as $P$'s one, should be univocally bound to something
that is a mere ``appearance to $P$''. Is it conceivable that a mere
``appearance to consciousness'' be the bearer of consciousness?\par

 To investigate this question let it first be reiterated that the
operationalist conception differs from pure idealism. In it, it is
assumed that some fundamental reality exists in an ontological sense;
that it is basically unknowable; and that it is somehow endowed with
(hidden) structures. And it is further postulated that the existence of
these structures is what accounts for the regularities we observe
within the phenomenal realm and synthesize in the form of rules
enabling us to predict our future observations. (Whether or not these
rules vaguely reveal us something concerning the said hidden structures
is an open question that lies outside our present subject).
Consequently, we have to do with two notions of an ontological nature,
the said, hidden mind-independent reality and consciousnesses, alias
minds (although, of course the latter may be taken to just be
components, or emanations, of the former).\par

 Now, normally intersubjective agreement holds between individual minds about what
they see (``what they see'' being an abbreviation for ``what they have the
impression of seeing'', for remember that objects are but appearances).
Physicists had to account for it and the way they managed to do so was
to invent universal predictive rules, at present remarkably synthesized
in the form of the {\it quantum} predictive rules. Now, the important point
concerning the question in hand is the (already mentioned) one that the
said rules (mainly the Born one) intersubjectively predict the
appearance (to the minds) of (phenomenal) macroscopic objects. And we
may consider this observational prediction to be experimentally
corroborated since we do see macroscopic objects. In particular, each
individual mind has the impression of being associated to one
particular such macroscopic object, called its own ``body'', as well as
that of seeing other similar objects. Now, according to the conception
in hand, for a long -- indeed a very long! -- time the fact that the
realist language is by far the most practical one (without it we could
hardly communicate as Bohr stressed) misled us into considering such
bodies to be ontologically real. And we even took them to be the
bearers, or supports, of our minds ... which nowadays directly leads to
the above stated difficulty (how could what is just an appearance, or
image, in our mind be the support, or seat, of a mind?). But within the
operationalist conception here under study bodies, which are mere
phenomena, are not in the least the supports of minds. Quite on the
contrary, the fact that physics is basically a set of observational
predictive rules indicates that the objects -- human bodies included --
essentially are intersubjective appearances to minds. We must then
consider it as a given fact that each individual mind has the
impression of being associated to one particular such macroscopic
object, called its own ``body'', and also has the impression of seeing
other similar objects, bodies included.\par

 Under these conditions, it is clear that the reasoning made near the end of Section~2 (the one
bearing on a ``measurement in the generalized sense'') holds good. At
time $t$ both physicist $P$ and her friend $F$ have the impression of seeing,
along with the instrument, two human bodies; and also have the
impression that one of these bodies is their own. And, moreover, $F$ has
the impression of seeing the pointer at one definite place on the
scale. When ensemble $E$ is considered the same holds true concerning all
the ``friends'' in $E$ and their respective pointers. According to standard
quantum mechanics, from this knowledge they have, each of them infers
definite probabilities concerning what results would be obtained in the
future if this or that measurement were done. And, again all these
possible observational predictions may -- theoretically -- be combined
according to classical and standard probability rules, so as to yield
the probabilities with which $P$ should herself obtain these results.
Now, to repeat, there are (environment involving) physical quantities
concerning which these probabilities do not coincide with those $P$
directly obtains by means of a quantum-mechanical calculation based on
the content of her own state of consciousness, {\it without} assuming the
friends to be in definite consciousness states. \par

Hence, to sum up, we found: \par

(i) That within the operational conception the received view
according to which bodies are (ontologically) the bearers of minds is
inconsistent and must be dropped\footnote{It is well known that, in the line of Kant's views, many
philosophers split the ``mind'' notion into two. They consider, on the
one hand a disembodied, non-personal so called ``epistemic subject'' or
``first person subject'', to whom objects, bodies etc. appear, and on the
other hand ``third person subjects'', who are minds as emanating from
bodies. We consider that the existence of the WignerÕs friend paradox 
seriously invalidates this conception and we strive here to avoid
it.}.\par

(ii) That to drop the said view
changes nothing to the fact that, as long as observational outcomes are
considered to be predictive, the Wigner's friend riddle remains
unsolved.\par

 But then the considerations that were put forward in Section
3 obviously yield the solution. Here, just as there, it suffices to
introduce the notion of ``internal states of consciousness'', taken to be
elements of a basic reality, and to assume that they are not publicly
accessible in general. The friends $F$ do really have the impression of
seeing this or that, but this does not necessarily count as public
knowledge. Hence, when $P$ calculates her probabilities of getting such
and such outcomes were she to perform such and such measurements
(possibly involving the $F$'s as objects), she should, as a matter of
principle, ignore the very existence of these private impressions the
$F$'s have. As for the $F$'s, however, due to the fact that they feel
themselves associated with macroscopic bodies it turns out that, in
practice, they may use the private impressions in question in order to
predict what they themselves will see.\par

 Note moreover that, since $P$ thinks of the $F$'s as being so associated, when directly calculating the
above-mentioned probabilities she must, concerning all practically
feasible measurements, take decoherence theory into account. This
implies that if she chose to violate the above-stated prescription and
take the existence of the said private impressions of the $F$'s into
account, this choice, concerning the said measurements, would by no
means lead her astray. She would make no detectable error as to the
probabilities of their outcomes. This explains that, not only in our
daily life but also in our scientific activity, we may do as if our
states of consciousness relative to factual data were totally
predictive (though, quite strictly speaking, we know they are not) and
as if those of our colleagues yielded predictions fully consistent with
our owns.\par \vskip 1 truecm

\newpage
\noi {\bf 5. THE CONSTRUCTION OF TIME}\par
\vskip 5 truemm

   When a physicist speculates his guidelines
include taking account of all relevant data and having strict regard
for consistency requirements. They do not specify that commonsense and
received views should be obeyed, be it only for the reason that quantum
mechanics gave him abundant proofs of the frailty of such handrails.
This remark, obviously, applies first of all to the ``space'' and
``object'' notions. While, according to commonsense corpuscles clearly
are {\it somewhere in} space, quantum physicists take them to be at no
definite place until we look. And, what is more, nonseparability
suggests that space itself should be but an ``a priori mode of our
sensibility'', as Kant thought. But then, what about time? In contrast
with space, up to now time resisted such an ``idealization'' process.
Attempts at making it a quantum observable were, on the whole,
unsuccessful. They were so, however, merely for technical reasons.
Within a speculative theory such as the present one, in which,
conceptually, minds are taken to be prior to objects and their
location, it is therefore quite natural -- be it only for curiosity
sake! -- that we should wonder whether, by any chance, they might be
conceptually prior to time as well. \par

Well, it turns out that, in fact,
it is possible to modify the foregoing model so as to make it
compatible with such a view. For this purpose, let us start with the
following set of ideas. Minds possess ontological existence, have
definite impressions, may communicate with one another and observe that
they agree about most impressions they have. Among these impressions
there is the one of space and of perceiving objects of various sizes
lying in space. There also is the impression of perceiving events
taking place in that space, as well as that of perceiving two events as
taking places either simultaneously or successively. Minds, moreover,
are able to imagine events they don't actually see and separate them in
two classes, those they ``believe in'' and call ``objective'' and those
that they call ``purely imaginary''. Let it be added that minds
primitively have the notions of ``before'' and ``after'' as well as the
ones of longer, shorter or equal duration between two pairs of
successive events. They also realize that if the duration between
events $a$ and $b$ is strictly longer than the one between events $c$ and $d$
they are able to think of an event $d'$ taking place after $a$ and before $b$
and such that the duration between $a$ and $d'$ is equal to that between $c$
and $d$. Now, it can be shown (Montbrial \cite{13r}, Kranz et al \cite{14r} as
quoted by Montbrial) that the fulfillment of this set of conditions
(plus a few others, needed for mathematical strictness but trivial in
this context) entails the existence of a mapping of the set of all the
objective events onto the set of real numbers. This mapping, then, is
just (intersubjective) time.\par

 Now, with the ``time'' notion at our disposal, we may essentially take over the reasoning developed towards
the end of the foregoing section. Indeed, we may argue that the minds
observe some events to regularly follow other ones. They take note of
such regularities and quite naturally (let us admit they have a natural
propensity to induction) they make use of them and build up
observational predictive rules, enabling them to calculate
probabilities of future events when past ones are known. In this way
they eventually construct the, classical and quantum, observational
predictive rules (now, in fact, unified since we know the former ones
to be deducible from the latter). With the help of these rules they
predict future instances of intersubjective agreement between them all.
And all the rest of our Section 4 construction may then be taken over
without change. Finally we therefore get a grand, somewhat
Plotinus-like, model, according to which Ultimate Reality generates or
contains (appropriate words are lacking!) minds that, just as Ultimate
Reality itself, are prior to both space and time. But these minds have
a great many impressions of all sorts, including the one of having
bodies, of being in time (as we just saw), of living in a Universe
endowed with events and laws, and so on. \par

This model is rational and -- as it seems to me -- truly consistent with what we know about basic
physics. In view of the great amount of questioning that, since its
advent, quantum mechanics raised with regard to its interpretation this
is certainly a nontrivial point, and if the model is to be taken
seriously I think it should be on this ground. Indeed, it cannot be on
any other one since it is at variance with basic things we think we
know!  Hence it may be taken into consideration only provided we
balance acceptation of it with due recognition of the {\it empirical reality}
notion, and acknowledge the central role of the latter in, not only our
activity but also our way of thinking. Empirical reality is the whole
set of the phenomena. Otherwise said, it is an understandable and
manageable mapping of everything that minds actually perceive and act
on. So that in practice -- in our daily life but also in pure science
(interpretation of quantum mechanics is no real exception for strictly
speaking it lies beyond science) -- we must do and think as if the model
it constitutes did represent Reality-per-se.  Concerning minds (alias
consciousness) this, in particular, implies that no changes at all are
necessary concerning the investigation procedures of neurologists.
True, these procedures are grounded on the view that minds are on the
dependence of bodies and more or less produced by them, which is quite
at variance with the views propounded here and in the foregoing
section. But such notions of dependence and causation are themselves of
a purely phenomenal nature, and this is what, in a phenomenal World,
makes them useful. They synthesize in a simple way some features of the
observational predictive rules that the minds discovered, so that,
clearly, we have to go on using them.\par \vskip 1 truecm

\noi {\bf 6. CONCLUSION}\par \vskip 5 truemm

The main result reached in this article is that
decoherence theory alone does not remove the Wigner's friend problem
but that the said problem can still be solved, by introducing the
``internal state of consciousness'' notion in the above described manner.
 More precisely, we found that this can be done in two ways. Admittedly
both of them are quite obviously at variance with the two most common
received views.  This however is partly due to the fact that, upon
inspection, the said views turn out to be inconsistent with one
another. It is impossible to claim at the same time that the meaning of
statements about the existence of objects of any kind (neurons
included) {\it boils down} to their method of verification and that neurons
are conceptually prior to the verifying agent. Looking for an approach
to the Wigner's friend problem made it necessary to clearly face this
conceptual difficulty. And finally two solutions were put forward.\par

One of them consists in openly accepting the idea, discarded both by the
quantum mechanics founding fathers and by the logical positivists, that
the theory should be ontologically interpreted. This solution is
self-consistent. It has the virtue of being compatible with the
intuitive views of both the ``man in the street'' and the scientists at
large, inasmuch as most of the latter take minds to emerge from matter.
It may however be criticized on the basis that it is vague without
being deliberately so. As a matter of principle, an ontologically
interpretable theory should describe reality ``as it really is'' and such
descriptions should match what is observed. The B.B. model does not
fulfill this condition and there are good reasons for believing
(nonseparability foremost) that other models yet to be invented cannot
fare better in this respect. \par

Conceptually, the other solution we found
is the opposite of this one. The Ultimate Reality it refers to is in
principle unknowable and is not what science deals with. Indeed, the
subject matter of the latter is just communicable human experience. In
other words it is the set of all the impressions human minds may have
and communicate to others. There is no doubt that the initiators of
this conception intended to place scientific knowledge on secure
grounds by setting it apart from ill-defined problems and particularly
those of an ontological nature. Quantum mechanics essentially developed
along these lines and the difficulties realist models steadily meet
with indicate that, indeed, they presumably constituted, for it, the
most favorable conceptual framework.  On the other hand, it is
unquestionable that the conception in question implicitly sets minds in
a privileged position, especially since not only matter but also space
and time are, in it, elements of experience. It was therefore to be expected
that its progressive development should,  at some time or other,
involve -- not just implicitly but explicitly -- the somewhat disturbing
idea that minds are conceptually prior to matter, space and time.\par

Anyhow, it may be difficult to choose between the two solutions we
found but it is worth stressing that the proto-mentality idea is
independent of this choice since we showed it to be consistent within
both conceptions.\par

Throughout Antiquity and the Middle Ages thinkers focused on purely
existential questions. But during the last centuries, and in relation
with the development of science, they were led to attach more and more
importance, even within the realm of pure knowledge, to what can be
achieved and to how it can be achieved. Maybe it is now time that
between these two tendencies -- existential and operational -- some
balance should be reached, and that the research tools of the latter
should be applied to the former. The present article may be viewed as a
step in this direction.

\end{document}